\documentclass[%
 twocolumn,
 amsmath,
 amssymb,
 aps,
 physrev,
]{revtex4-2}

\usepackage{graphicx}
\usepackage{dcolumn}
\usepackage{bm}
\usepackage[pdftex,table]{xcolor}
\usepackage{hyperref}
\usepackage{enumitem}
\usepackage[colorinlistoftodos]{todonotes}

\hypersetup{
  colorlinks=true,
  citecolor=blue,
  linkcolor=blue,
  urlcolor=blue
}

\newcommand{\Lam}{\Lambda}
\newcommand{\Gs}{G_{i}^{\mathrm{Strips}}}
\newcommand{\Fp}{F_{i}^{\mathrm{Pixels}}}
\newcommand{\taup}{\tau^{\prime}}

\newcommand{\closureIonRange}{11\%}

\newcommand{\mExpOne}{910}

\newcommand{\mObsFifty}{740}

\newcommand{\mObsLoOne}{140}

\newcommand{\mObsNicThreeOne}{1030}

\newcommand{\mObsNicThreeThirtyK}{350}

\newcommand{\mObsNicTwoOne}{970}

\newcommand{\mObsNicTwoSeventy}{730}
\newcommand{\mObsNicTwoTenK}{290}

\newcommand{\mObsNicTwoThirtyK}{320}

\newcommand{\mObsOne}{870}

\newcommand{\mObsSeventy}{620}
\newcommand{\mObsTenK}{260}
\newcommand{\mObsThirty}{830}
\newcommand{\mObsThirtyK}{280}

\newcommand{\mObsTwoK}{130}
\newcommand{\mSingletOne}{990}
\newcommand{\mSingletSeventy}{720}
\newcommand{\mSingletTenK}{280}

\newcommand{\mSingletThirtyK}{300}
\newcommand{\noisePix}{40\%}
\newcommand{\noiseStrip}{15\%}
\newcommand{\presHundredSix}{0.07}
\newcommand{\presOneSix}{0.40}
\newcommand{\presThirtySix}{0.32}
\newcommand{\resultSummary}{masses up to 870 (970)\,GeV are excluded for $\Lam\lesssim70$\,eV, where the quirks behave as independent heavy stable charged particles. For $\Lam\gtrsim10$\,keV, where the pair moves as one doubly ionizing object, the limit is 280 (320)\,GeV. In between, the quirks turn around inside the detector and the search has no sensitivity}
\newcommand{\singletSummary}{For the $SU(2)$-singlet benchmark of the FASER search, with two infracolors, the limits reach 990\,GeV below 100\,eV, where FASER has no sensitivity, and 300\,GeV above 10\,keV, beyond the FASER exclusion of 175\,GeV}

\newcommand{\trigHundredSix}{0.31}

\newcommand{\trigThirtySix}{0.48}
\newcommand{\xsecOneFourHundred}{$7.2\times10^{-6}$}
\newcommand{\xsecTwoHundred}{$0.126$}

\begin{document}

\preprint{APS/}

\title{\textbf{Constraints on quirks from a Geant4-based reinterpretation of the CMS search for heavy stable charged particles}}

\author{Tamas Almos Vami}
\email{tamasvami@ucsb.edu}
\affiliation{University of California Santa Barbara, Santa Barbara, CA, USA}

\date{\today}

\begin{abstract}
Quirks are new heavy fermions with electroweak charges that are also bound by a confining force whose scale $\Lam$ is far below their mass. A produced pair therefore stays connected by a macroscopic string. We reinterpret the CMS search for heavy stable charged particles in 101\,fb$^{-1}$ of 13\,TeV proton-proton collisions for Drell-Yan produced quirks. The quirk dynamics is simulated with Geant4 in a parametric CMS-like detector, and the event selection is emulated and combined with the public likelihoods. The emulation reproduces the published limits on a heavy charged lepton within \closureIonRange. For an $SU(2)$-doublet quirk with one (two) infracolors, \resultSummary. \singletSummary.
\end{abstract}

\maketitle

\section{Introduction}
\label{sec:intro}

Quirks~\cite{Okun:1980kw,Kang:2008ea} are new heavy fermions $Q$ with electroweak charges that are also charged under a confining gauge group, infracolor, whose confinement scale $\Lam$ is much smaller than the quirk mass $m_Q$, as in folded supersymmetry~\cite{Burdman:2008ek}. Since no infracolored particle is lighter than $\Lam$, the infracolor flux tube between a pair of quirks cannot break: a quirk-antiquirk pair produced at a collider stays connected by a string of tension $\sigma \simeq \Lam^{2}$ and oscillates around its center of mass until it annihilates. The amplitude of the oscillation is set by the kinetic energy of the quirks in the rest frame of the pair, $d_{\max}\simeq 2(E^{*}-m_Q)/\Lam^{2}$, where $E^{*}$ is the energy of each quirk in that frame. For $E^{*}-m_Q$ of order 100\,GeV, the pair is meters wide at $\Lam\simeq100$\,eV, centimeters wide at 1\,keV, and narrower than the pitch of a silicon detector at a few tens of keV (Table~\ref{tab:amp}).

\begin{table}[h!]
\caption{String tension and largest separation of a quirk pair with $E^{*}-m_Q=100$\,GeV, and the resulting detector signature.}
\label{tab:amp}
\begin{ruledtabular}
\begin{tabular}{rrrl}
$\Lam$ & $\Lam^{2}/\hbar c$ [GeV/m] & $d_{\max}$ & Signature \\
\hline
1\,eV & 0.005 & 40\,km & independent HSCPs \\
30\,eV & 4.6 & 44\,m & HSCPs, deflected tracks \\
100\,eV & 51 & 4\,m & turn in the calorimeters \\
300\,eV & 460 & 44\,cm & turn in the tracker \\
1\,keV & $5\times10^{3}$ & 4\,cm & cm-wide pair \\
3\,keV & $5\times10^{4}$ & 4\,mm & mm-wide pair \\
10\,keV & $5\times10^{5}$ & 0.4\,mm & partly merged \\
30\,keV & $5\times10^{6}$ & 44\,$\mu$m & merged, $2\times$ ionization \\
\end{tabular}
\end{ruledtabular}
\end{table}

At a general-purpose Large Hadron Collider (LHC) detector, quirks therefore appear as heavy stable charged particles (HSCPs) for $\Lam$ of a few eV, as pairs that turn around inside the detector for $\Lam$ between about 100\,eV and 1\,keV, and as one neutral object with twice the ionization of a single charge above a few keV. Reinterpretations of HSCP and monojet searches constrain $\Lam$ of about 1 to 100\,eV~\cite{Farina:2017ywr}, the D0 experiment searched for quirks with $\Lam$ from 10\,keV to 1\,MeV~\cite{D0:2010kkd}, and dedicated strategies target slow pairs whose tracks do not bend~\cite{Knapen:2017kly}, the ionization of quirks in the tracker~\cite{Li:2019wce}, quirks that stop in the detector~\cite{Evans:2018jmd}, and the LHCb~\cite{CidVidal:2020okg} and FASER~\cite{Li:2021tsy,FASER:quirkpheno,FASER:quirksearch} detectors.

In this paper, we derive constraints on Drell-Yan (DY) produced quirks from the CMS search for HSCPs with large ionization energy loss in 2017 and 2018 data~\cite{CMS:EXO18002,Vami:2025jspc}, using only public information: the HEPData~\cite{Maguire:2017ypu} record of the search~\cite{CMS:EXO18002:hepdata}, which contains the likelihoods (statistical models in the Combine format~\cite{CMS:2024onh}), the selection efficiencies, and the trigger efficiency maps. The quirk propagation is simulated with a standalone Geant4~\cite{GEANT4:2002zbu,Allison:2006ve,Allison:2016lfl} application in a parametric CMS-like detector. The detector response and the event selection are emulated, and the emulation is calibrated and validated by reproducing the published limits on a heavy charged lepton $\taup$ with charge $|Q|=1e$.

\section{Quirk production and dynamics}
\label{sec:model}

We consider a vector-like quirk with the electroweak quantum numbers of a fourth-generation lepton doublet partner (a $\taup$ with $|Q|=1e$, hypercharge as in the Standard Model), produced in pairs through $q\bar{q}\to\gamma^{*}/Z\to Q\bar{Q}$. Events are generated with \textsc{Pythia}~8.3~\cite{Bierlich:2022pfr} at $\sqrt{s}=13$\,TeV with the CP5 tune~\cite{CMS:2018tau} and the NNPDF3.1 parton distributions~\cite{NNPDF:2017mvq}, using the leading order (LO) cross section; with $N_{\mathrm{IC}}$ infracolors the cross section is $N_{\mathrm{IC}}$ times larger and the kinematics are unchanged. The cross section falls from \xsecTwoHundred\,pb at $m_Q=200$\,GeV to \xsecOneFourHundred\,pb at 1.4\,TeV.

Except at the lowest $\Lam$, the string tension (Table~\ref{tab:amp}) dominates every other force in the detector, already forty times the magnetic force on a relativistic unit charge in 3.8\,T at $\Lam=100$\,eV, so the quirk transport has to be solved together with the interactions with matter.

\section{Simulation}
\label{sec:sim}

The quirk propagation uses \texttt{quirksim} (Appendix~\ref{sec:code}), a standalone Geant4 application built on the same quirk transportation that the FASER search used for its detector simulation~\cite{FASER:quirksearch}. Each quirk is transported in short steps under the combined Lorentz and string forces, in alternation with its partner: a quirk is suspended after each step and resumed once its partner has caught up, so that the string force is always computed from the partner trajectory at the same time. Ionization, multiple scattering, and bremsstrahlung of the quirks are simulated with the standard electromagnetic models of Geant4 version 11.4.1. Being color neutral, the quirks have no hadronic interactions.

The detector is a set of cylinders and disks with the radii, thicknesses, materials, and magnetic field of the CMS detector~\cite{CMS:2008xjf}, from the pixel detector to the muon stations, with 3.8\,T inside the solenoid and 1.8\,T in the return yoke. The double-sided strip modules of the first two inner and outer barrel layers are modeled as one sensor of twice the thickness. Varying the tracker material by $\pm50$\% or the calorimeter absorber by $-20$\% changes the acceptance observables of quirk pairs by less than the statistical precision of 200-event samples.

Only the quirks are passed to Geant4, and only events that can pass the selection, as determined at the generator level, are simulated; the fraction kept enters the efficiency.

\section{Emulated reconstruction and selection}
\label{sec:emulation}

The Geant4 hits are turned into clusters: hits of the two quirks within twice the pitch in a sensor form one cluster with the summed energy and the energy-weighted position, and hits more than 25\,ns later than a particle at the speed of light are discarded. A track candidate in which at least half of the clusters contain both quirks is treated as one merged track of the pair. Each candidate is fitted with a circle in the transverse plane and a line in $(r,z)$, with the CMS hit resolutions and a multiple scattering term, which gives $p_{\mathrm{T}}$, $\chi^{2}/\mathrm{dof}$, the impact parameters $d_{xy}$ and $d_z$, and $\sigma_{p_{\mathrm{T}}}/p_{\mathrm{T}}^{2}$.

The energy loss per unit path length $dE/dx$ of each cluster is smeared by an effective resolution and compared to the distribution of a minimum ionizing particle (MIP), obtained in the same detector from charged particles at $\beta\gamma=70$ with the same smearing, separately for each sensor thickness. The probability $P_i$ that a MIP gives a smaller energy loss than cluster $i$ enters the two discriminants of the search: the strip discriminant
\begin{equation}
\Gs = \frac{3}{N}\left[\frac{1}{12N} + \sum_{i=1}^{N} P_i\left(P_i - \frac{2i-1}{2N}\right)^{2}\right],
\end{equation}
with the strip clusters ordered by increasing $P_i$, and the pixel discriminant
\begin{equation}
\Fp = 1-\Pi\sum_{k=0}^{N-1}\frac{(-\ln\Pi)^{k}}{k!},
\end{equation}
with $\Pi=\prod_{i}(1-P_i)$ over the $N$ pixel clusters outside the innermost layer, which is one minus the combined probability that a MIP gives a larger energy loss than every cluster. The effective resolutions, \noisePix\ for pixels and \noiseStrip\ for strips, are the only tuned parameters of the emulation: they are fixed by requiring that the fraction of the $\taup$ signal in the $\Fp<0.9$ (FAIL) region and in the last $\Gs$ bin of the PASS region match the published likelihoods (Section~\ref{sec:closure}).

The relative mini-isolation and the track isolation are computed from the generator-level visible particles around the candidate direction, with the partner quirk added when it is inside the cone and not part of the candidate. The energy-to-momentum ratio adds the energy deposited by the quirks themselves in the calorimeters. The single-muon trigger is applied as an efficiency per quirk that reaches at least two muon stations, from the published maps for $\tilde{\tau}$ leptons as a function of $\beta$ in six $|\eta|$ bins~\cite{CMS:EXO18002:hepdata}. For quirks, $\beta$ is the effective velocity to the first muon station (distance from the origin over arrival time), which coincides with the true $\beta$ for a free HSCP to better than 1\% above $\beta=0.4$, and two quirks that arrive within $\Delta R<0.1$ count as one muon object. The up and down variations of the maps, which correspond to a 1.5\,ns shift of the muon arrival time, provide the trigger uncertainty.

The candidate must pass the preselection of the search: $p_{\mathrm{T}}>55$\,GeV, $|\eta|<1$, at least two pixel hits outside the innermost layer, more than 80\% valid hits, at least ten $dE/dx$ measurements, $\chi^{2}/\mathrm{dof}<5$, $d_z<0.1$\,cm, $d_{xy}<0.02$\,cm, mini-isolation below 0.02, track isolation below 15\,GeV, $E/p<0.3$, $\sigma_{p_{\mathrm{T}}}/p_{\mathrm{T}}^{2}<0.0008$\,GeV$^{-1}$, and $\Fp>0.3$. The signal region requires $p_{\mathrm{T}}>200$\,GeV and is binned in $\Gs$ (nine bins) separately for $\Fp>0.9$ (PASS) and $0.3<\Fp<0.9$ (FAIL).

The emulated signal yields replace the signal templates of the published likelihoods, which are otherwise used unchanged: a binned PASS/FAIL model in which the background is constrained by the FAIL region. The signal shape uncertainties of the published model are replaced by an overall uncertainty of 15\% from the closure test below, the trigger uncertainty, the MC statistical uncertainty of the emulated signal, and the 1.6\% integrated luminosity uncertainty. Points with fewer than three effective MC events in the signal region are not used. Upper limits at 95\% confidence level are computed with the CL$_{\mathrm{s}}$ criterion~\cite{Junk:1999kv,Read:2002hq} in the asymptotic approximation~\cite{Cowan:2010js} with a standalone build of Combine~\cite{CMS:2024onh}.

\section{Validation with a heavy charged lepton}
\label{sec:closure}

The $|Q|=1e$ $\taup$ of the CMS search is what a quirk becomes when the string is negligible, and its published likelihoods contain the signal templates, so it fixes both the calibration and the precision of the emulation. We generate it with the generator settings of the search (published with the HEPData record), simulate it with the same chain as the quirks with the string switched off in practice ($\Lam=1$\,eV, a string amplitude of tens of kilometers), and treat the published signal yields as the target.

\begin{figure}[h!]
\centering
\includegraphics[width=0.48\textwidth]{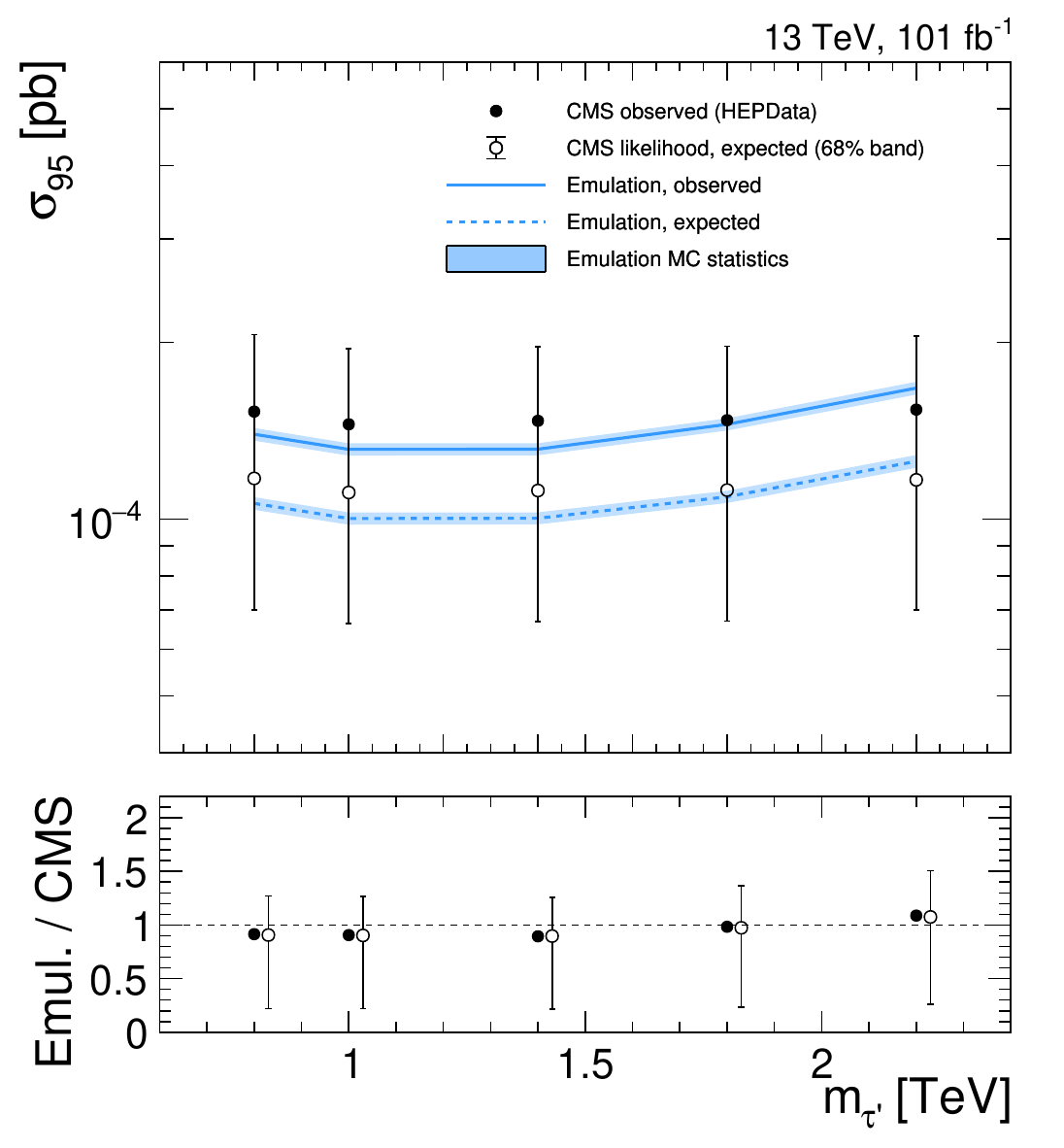}
\caption{Upper limits at 95\% confidence level on the cross section of a $|Q|=1e$ $\taup$: published observed (filled points) and expected (open points, with 68\% band) limits, and the limits with the emulated signal on the same background model (lines, band: MC statistical uncertainty). The lower panel shows the emulated-to-published ratio.}
\label{fig:closure}
\end{figure}

\begin{figure}[h!]
\centering
\includegraphics[width=0.48\textwidth]{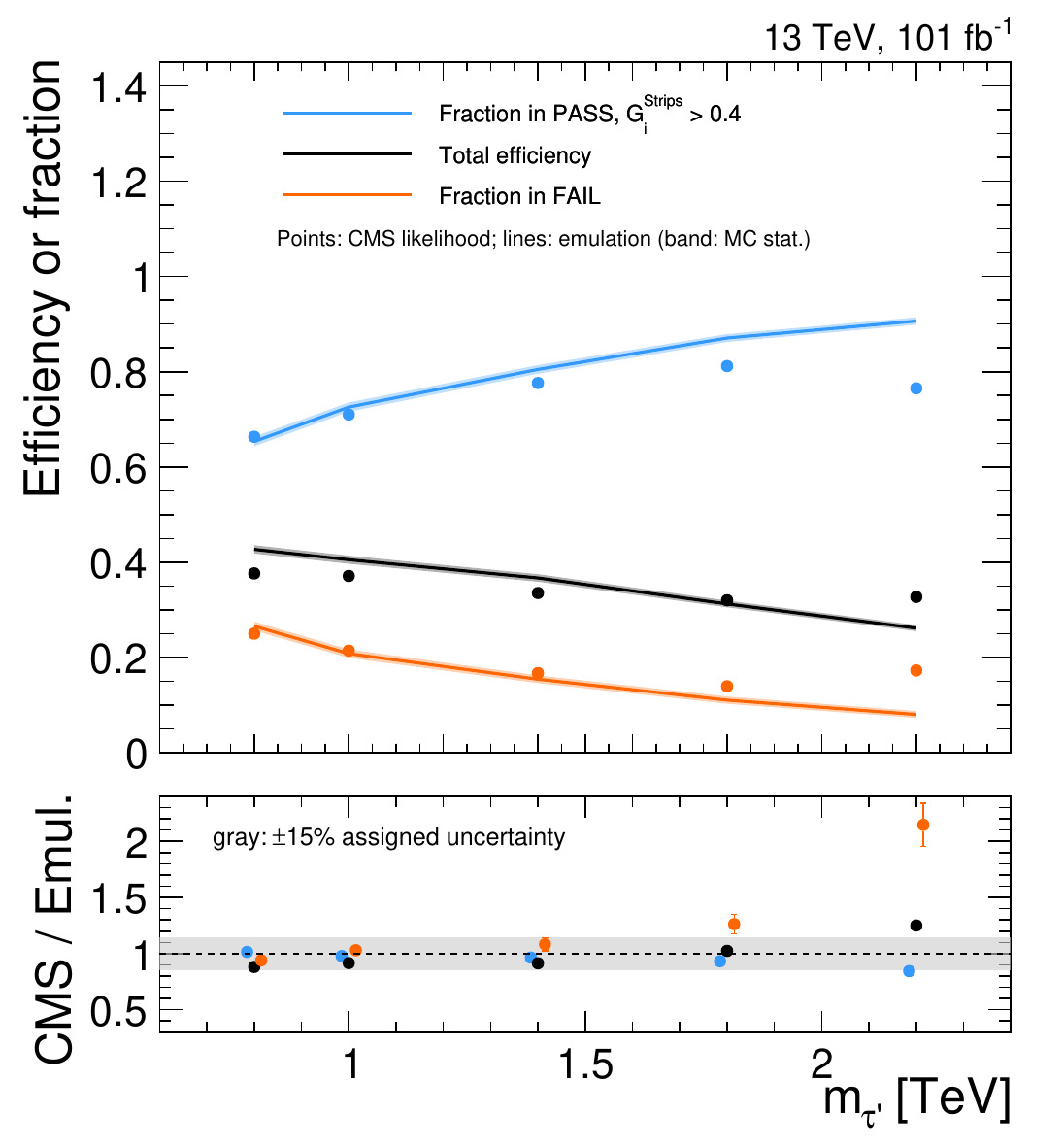}
\caption{Total signal efficiency of the $|Q|=1e$ $\taup$ and its fractions in the FAIL region and in the most sensitive bin, from the published likelihood (points) and the emulation (lines, band: MC statistical uncertainty). The lower panel shows the published-to-emulated ratio; the gray band is the 15\% uncertainty assigned to the emulated efficiency.}
\label{fig:closurefrac}
\end{figure}

Figure~\ref{fig:closure} compares the emulated and published 95\% confidence level upper limits on the $\taup$ cross section, both computed on the published background model. The emulated limits agree with the published ones within \closureIonRange\ from 0.8 to 2.2\,TeV, for both the observed and the expected limits. The total signal efficiency agrees within 13\% up to 1.8\,TeV, and the split between the FAIL and PASS regions and the fraction in the most sensitive bin ($\Fp>0.9$, $\Gs>0.4$) agree within 0.03 and 0.06 (Fig.~\ref{fig:closurefrac}). Above 1.8\,TeV the emulation underestimates the FAIL fraction, most likely because the response of the CMS pixel detector to very slow particles, such as late charge collection and readout saturation, is not modeled; this mass range lies beyond the quirk masses considered here, and the limits still agree within 9\%. Normalized to the events with a candidate within $|\eta|<1$, the published $\taup$ cutflow is reproduced step by step within 5\%: the full preselection keeps 92 to 97\% of these events in CMS and 97 to 99\% in the emulation, with the largest difference, 2 to 3\%, at the pixel-hit requirement, since the modeled pixel detector has no inactive regions. We assign a 15\% uncertainty to the emulated signal efficiency to cover these differences.

\section{Quirk signatures}
\label{sec:signatures}

\begin{figure*}[t]
\centering
\includegraphics[width=0.32\textwidth]{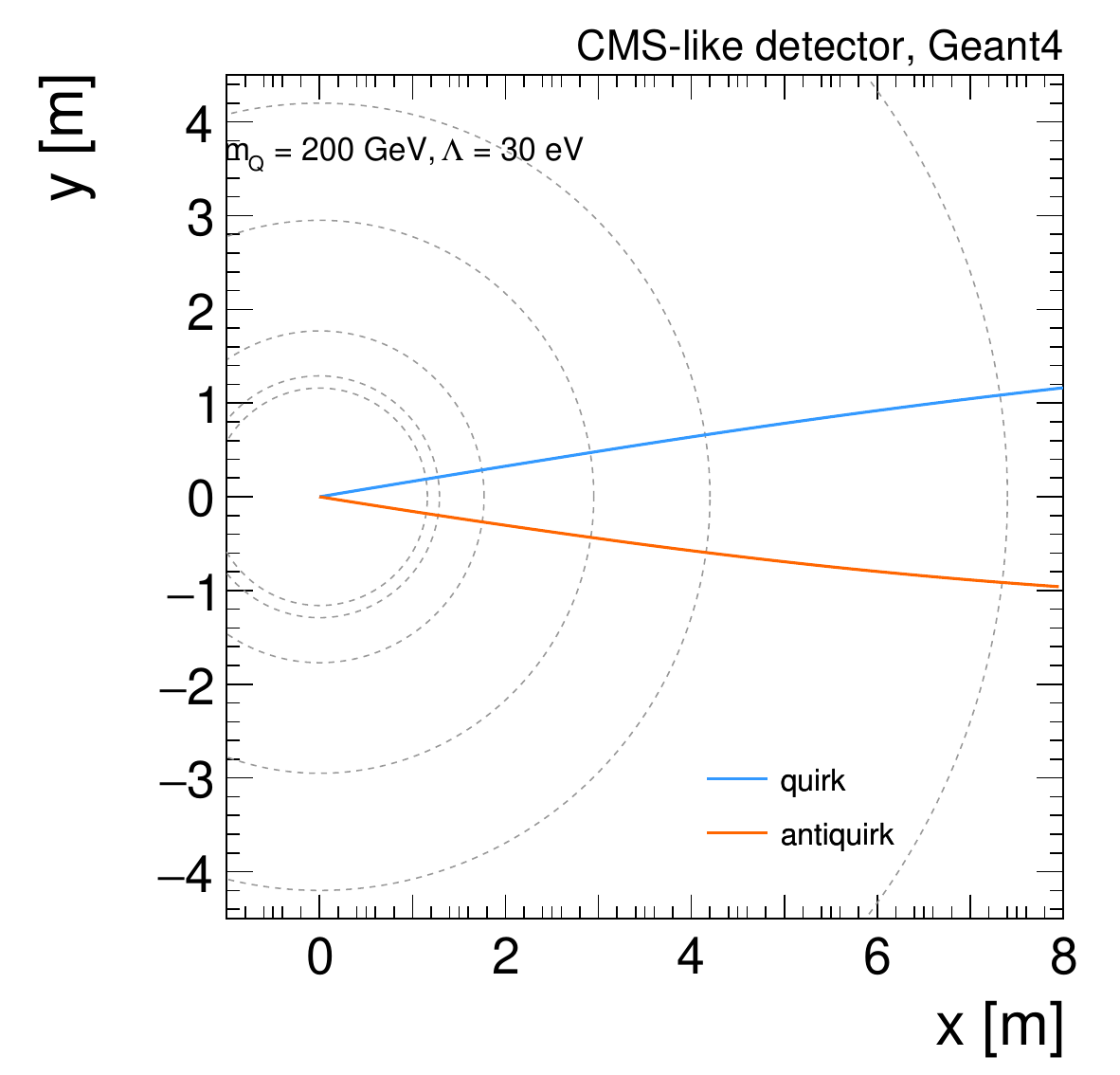}
\includegraphics[width=0.32\textwidth]{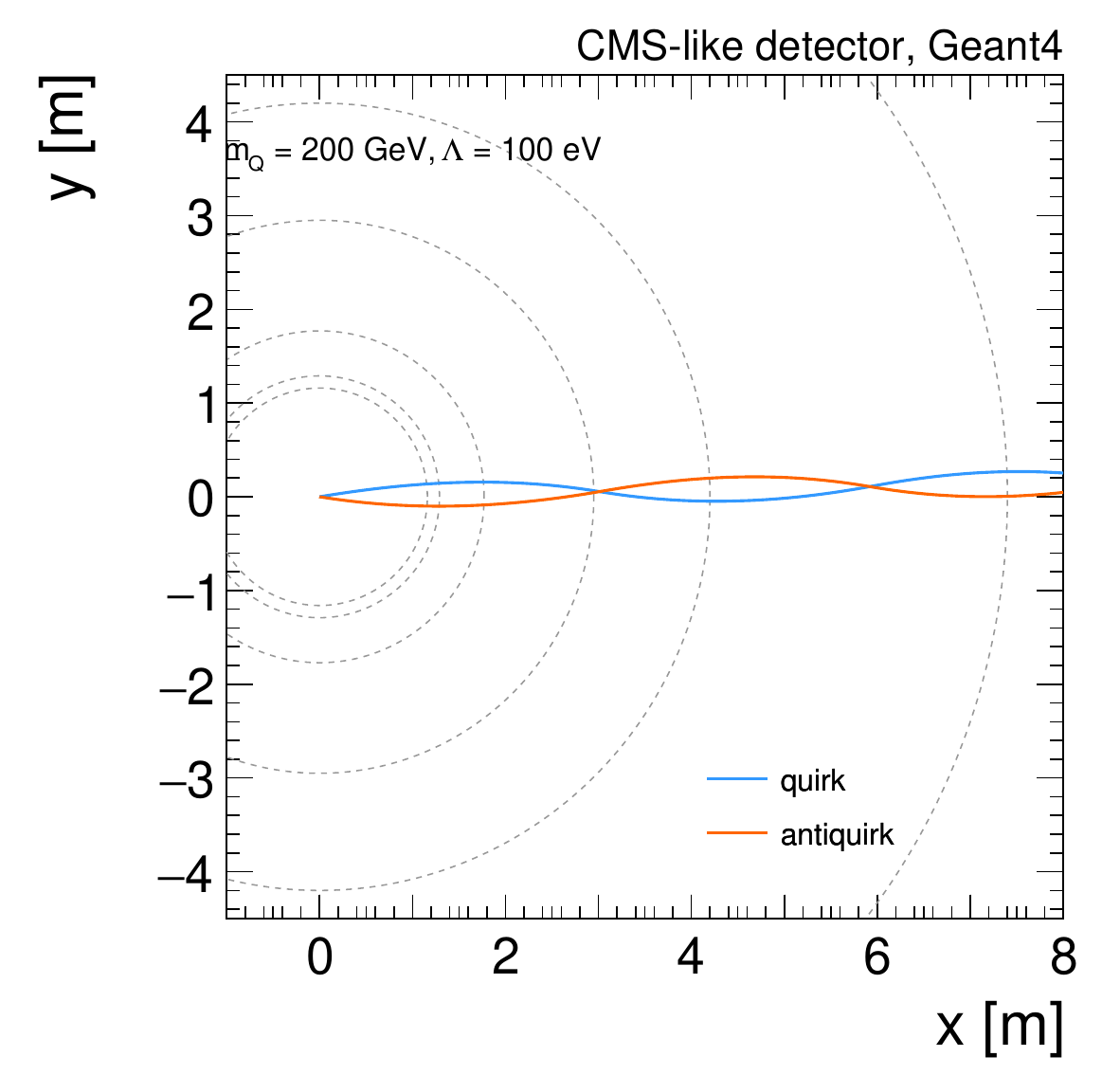}
\includegraphics[width=0.32\textwidth]{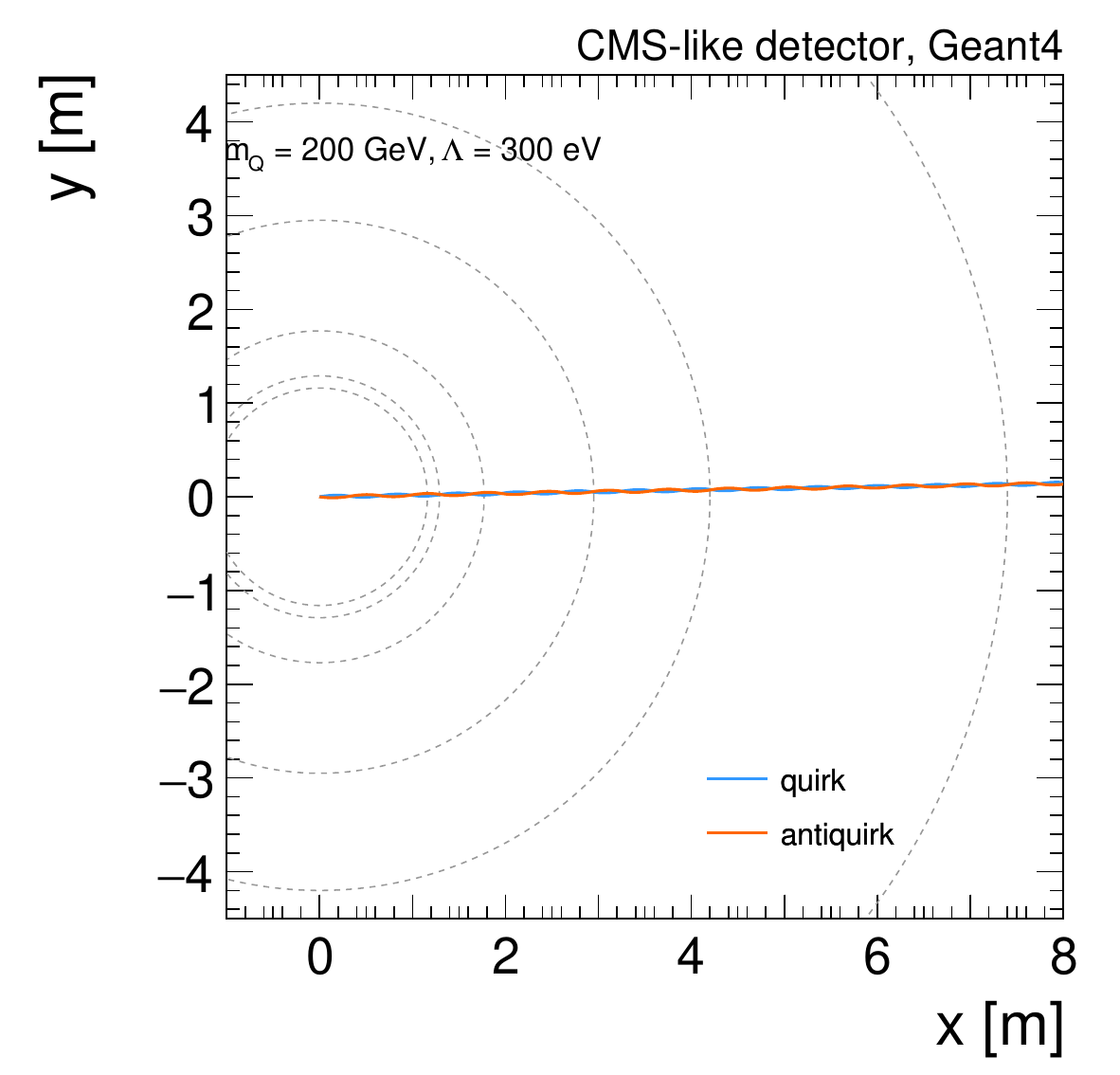}
\caption{Transverse view of the trajectories of a quirk pair with $m_Q=200$\,GeV and momenta $(300, 50, 100)$ and $(250, -40, 150)$\,GeV in the CMS-like detector at $\Lam=30$\,eV (left), 100\,eV (middle), and 300\,eV (right). The dashed circles mark the outer radii of the tracker, the electromagnetic and hadron calorimeters, the solenoid, and the muon system.}
\label{fig:display}
\end{figure*}

Table~\ref{tab:amp} summarizes the scales that decide the detector signature, and Figure~\ref{fig:display} shows the same quirk pair simulated at three values of $\Lam$.

\begin{figure}[h!]
\centering
\includegraphics[width=0.48\textwidth]{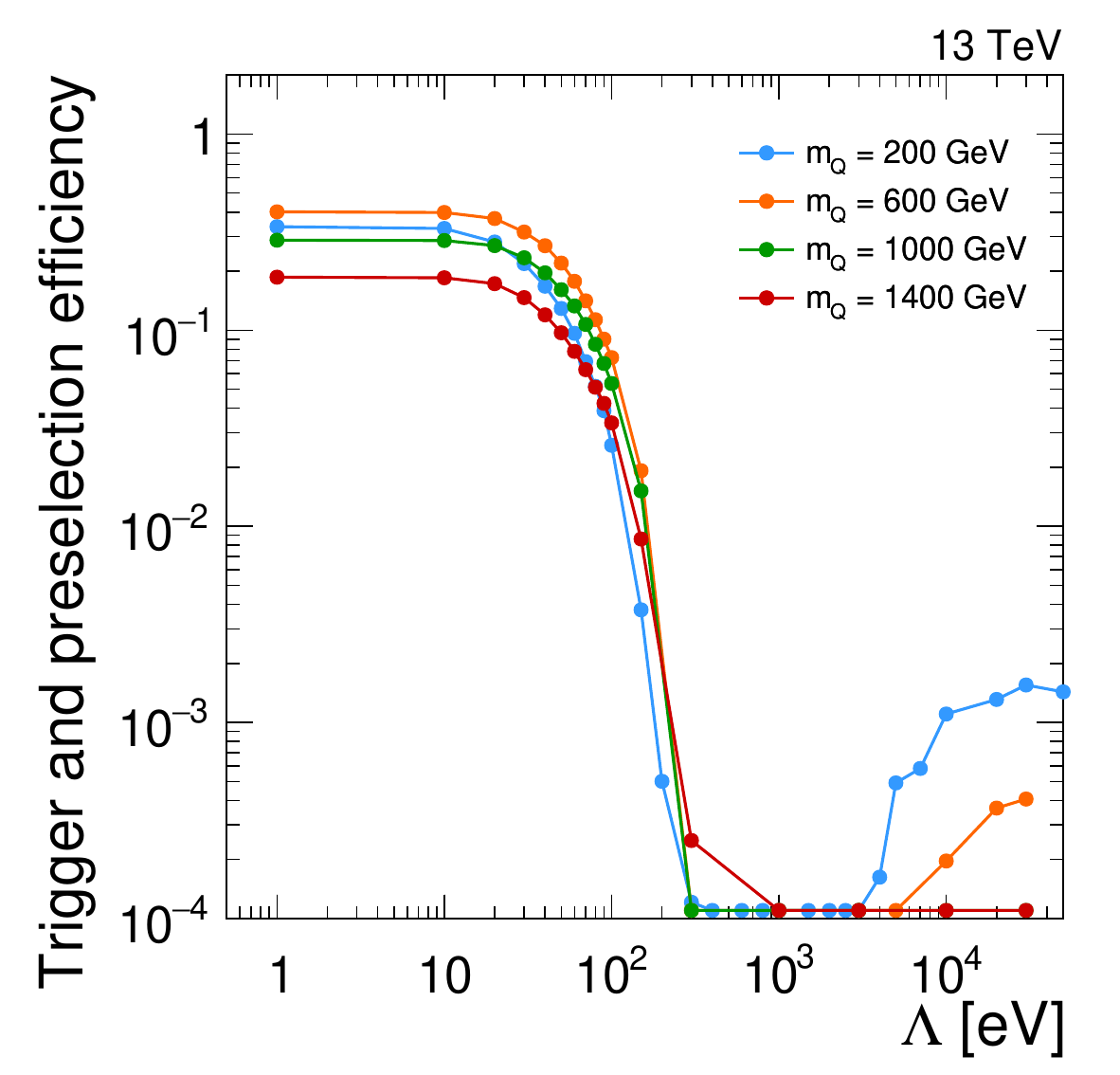}
\caption{Fraction of generated quirk pairs that pass the trigger and the preselection, as a function of $\Lam$ for four quirk masses.}
\label{fig:presel}
\end{figure}

Figure~\ref{fig:presel} shows how the trigger and preselection efficiency changes with $\Lam$. At $\Lam=1$\,eV, \presOneSix\ of the generated pairs with $m_Q=600$\,GeV pass the trigger and the preselection, a value close to that of the $\taup$ at similar mass; the remaining difference comes from the slower and more forward kinematics of the doublet quirk. The efficiency falls steeply between 30 and 100\,eV (\presThirtySix\ and \presHundredSix\ at 600\,GeV), long before the quirks turn around inside the tracker. The trigger is not the reason: the quirks still reach the muon system with a similar efficiency (\trigThirtySix\ and \trigHundredSix). The selection step that removes the events is the track $\chi^{2}/\mathrm{dof}<5$ requirement: already at $\Lam=30$\,eV the string tension, 4.6\,GeV/m, exceeds the magnetic force on a unit charge, and whenever the string is not aligned with the quirk momentum (a boosted pair, or a pair recoiling against initial-state radiation) it deflects the trajectory transversely by more than the hit resolution over the length of the tracker.

From 300\,eV to 1\,keV no event passes the preselection: the quirks are centimeters to tens of centimeters apart inside the tracker, so each leaves an oscillating sequence of hits that no helix fit accepts, while their clusters are too far apart to merge. From 1 to 3\,keV the lightest pairs, up to about 150\,GeV, start to pass as merged tracks, and at 10 and 30\,keV most selected candidates at all masses are merged tracks of the pair, with $\Gs$ close to one. Their acceptance is limited by the trigger, since the pair is heavy and slow ($\beta$ between 0.2 and 0.6 for most central pairs, where the muon trigger efficiency is small), and by $\chi^{2}/\mathrm{dof}<5$, since the energy-weighted cluster positions still follow the residual oscillation of the pair.

\section{Results}
\label{sec:results}

\begin{figure}[h!]
\centering
\includegraphics[width=0.48\textwidth]{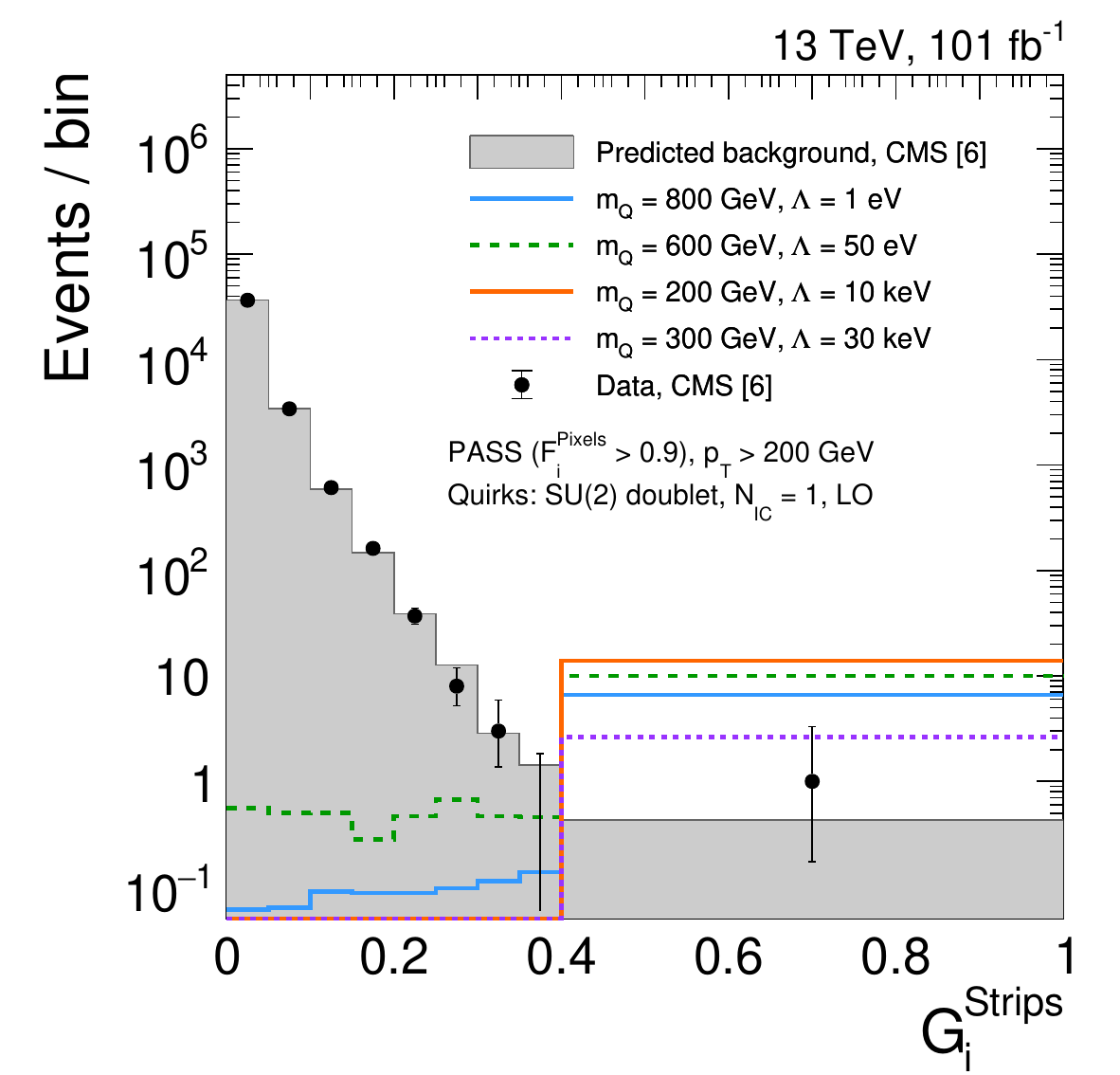}
\caption{Distribution of $\Gs$ in the PASS region ($\Fp>0.9$, $p_{\mathrm{T}}>200$\,GeV) of the CMS search: observed events (points) and predicted background (gray), from the HEPData record~\cite{CMS:EXO18002:hepdata}, with the emulated signal of an $SU(2)$-doublet quirk ($N_{\mathrm{IC}}=1$, LO cross section) for four values of $(m_Q,\Lam)$.}
\label{fig:gstrip}
\end{figure}

Figure~\ref{fig:gstrip} shows the $\Gs$ distribution of the PASS region ($p_{\mathrm{T}}>200$\,GeV) from the CMS search, observed and predicted, with the emulated signal of four quirk benchmarks. In every regime, the quirk signal is concentrated in the last bin, $\Gs>0.4$, where one event is observed for a predicted background of 0.43 events; the residual signal at lower $\Gs$ comes from fast quirks at small $\Lam$.

\begin{figure}[h!]
\centering
\includegraphics[width=0.48\textwidth]{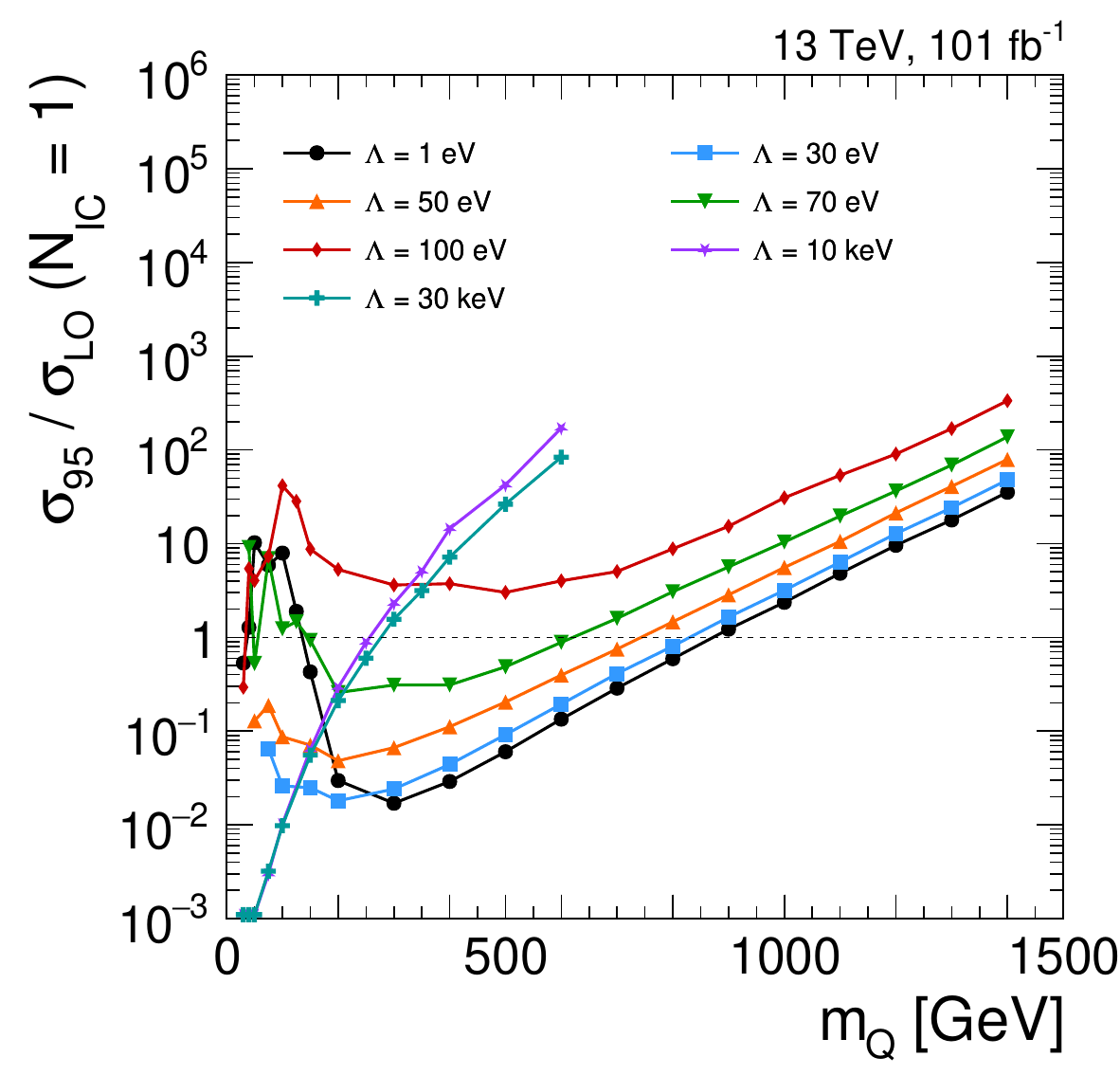}
\caption{Observed upper limit at 95\% confidence level on the DY quirk pair cross section, divided by the LO cross section of an $SU(2)$-doublet quirk with $N_{\mathrm{IC}}=1$, as a function of the quirk mass for several values of $\Lam$. Values below one are excluded; for $N_{\mathrm{IC}}$ infracolors the threshold is $N_{\mathrm{IC}}$.}
\label{fig:rlim}
\end{figure}

\begin{figure}[h!]
\centering
\includegraphics[width=0.48\textwidth]{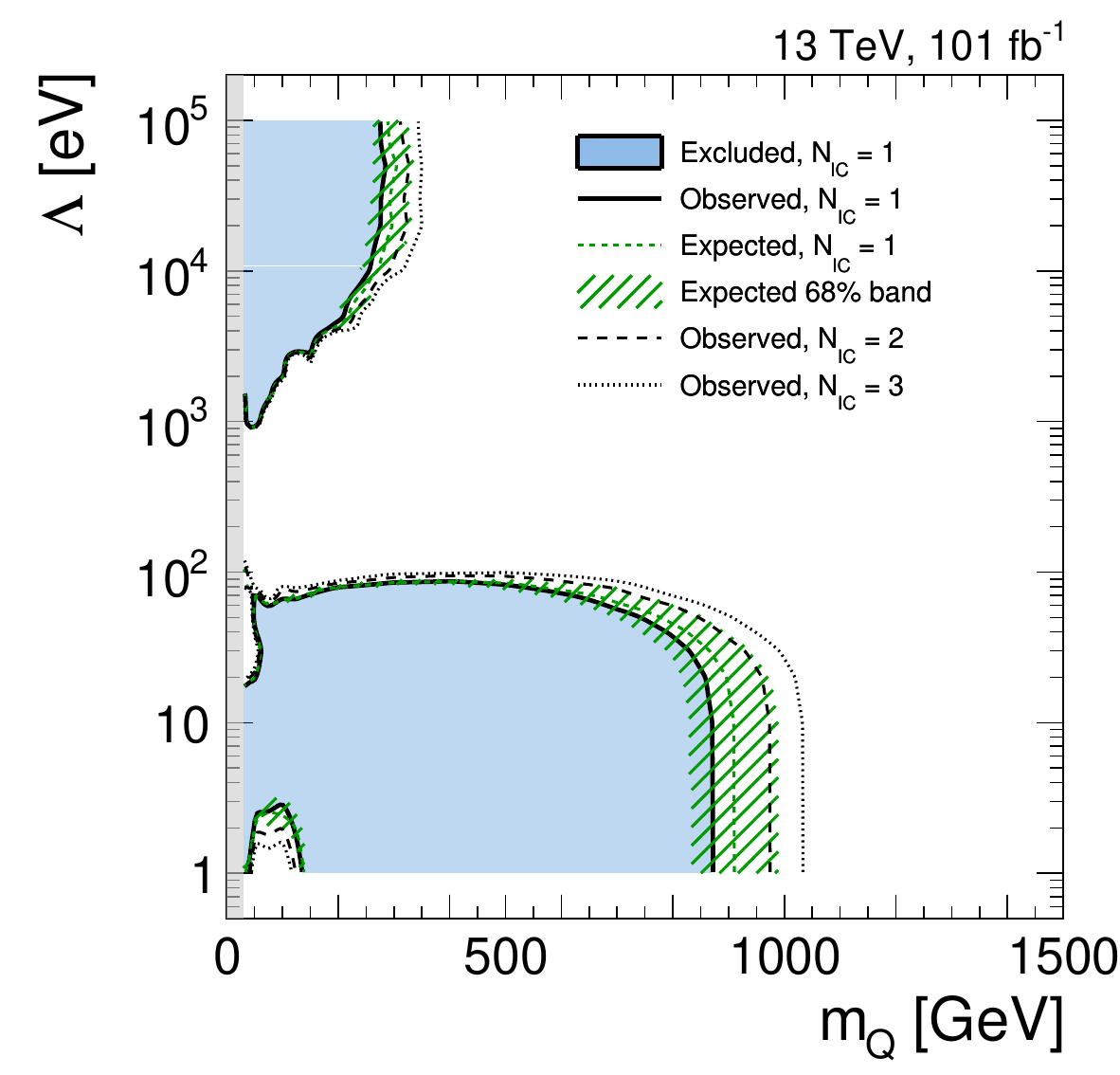}
\caption{Excluded region in the plane of the quirk mass and the confinement scale $\Lam$ (shaded: excluded for $N_{\mathrm{IC}}=1$; hatched: 68\% band of the expected exclusion; the gray region is not simulated): observed for $N_{\mathrm{IC}}=1$, 2, and 3 (black lines) and expected with its 68\% band for $N_{\mathrm{IC}}=1$.}
\label{fig:mlim}
\end{figure}

Figure~\ref{fig:rlim} shows the upper limit on the quirk pair cross section divided by the LO cross section for $N_{\mathrm{IC}}=1$, for each $\Lam$. Below about 150\,GeV the limits are not monotonic in mass: a free quirk with $p_{\mathrm{T}}>200$\,GeV becomes too fast ($\beta\approx0.97$ at 75\,GeV) to ionize much more than a minimum ionizing particle, while the cross section keeps rising, so that $\sigma_{95}/\sigma$ peaks between 50 and 100\,GeV, at a mass that depends on $\Lam$ and is affected by the limited MC statistics in this region. Figure~\ref{fig:mlim} translates these limits into the excluded region in the $(m_Q,\Lam)$ plane for $N_{\mathrm{IC}}=1$, 2, and 3, interpolating $\log(\sigma_{95}/\sigma)$ linearly in mass and in $\log\Lam$ between grid points and smoothing it mildly against the statistical scatter of the simulated points.

For $N_{\mathrm{IC}}=1$, quirk pairs are excluded from $m_Q=\mObsLoOne$ to \mObsOne\,GeV (expected \mExpOne\,GeV) for $\Lam=1$\,eV, where the quirks are free HSCPs, \mObsThirty\,GeV at 30\,eV, \mObsFifty\,GeV at 50\,eV, and \mObsSeventy\,GeV at 70\,eV. From 10 to 50\,eV the string force, 0.5\,GeV/m at 10\,eV, is comparable to the magnetic force on a slow quirk and can straighten its track, so that slow, highly ionizing quirks ($\beta\approx0.4$) are reconstructed with $p_{\mathrm{T}}>200$\,GeV and the exclusion reaches down to 50\,GeV. For $\Lam$ from 100 to 800\,eV, the search has no sensitivity above 40\,GeV. For $\Lam$ above 1\,keV, where the pair is increasingly reconstructed as one merged track, the excluded mass grows from \mObsTwoK\,GeV at 2\,keV to \mObsThirtyK\,GeV at 30\,keV. For $N_{\mathrm{IC}}=2$ (3), the limits at 1\,eV and 30\,keV rise to \mObsNicTwoOne\ (\mObsNicThreeOne) and \mObsNicTwoThirtyK\ (\mObsNicThreeThirtyK)\,GeV.

\begin{figure}[h!]
\centering
\includegraphics[width=0.48\textwidth]{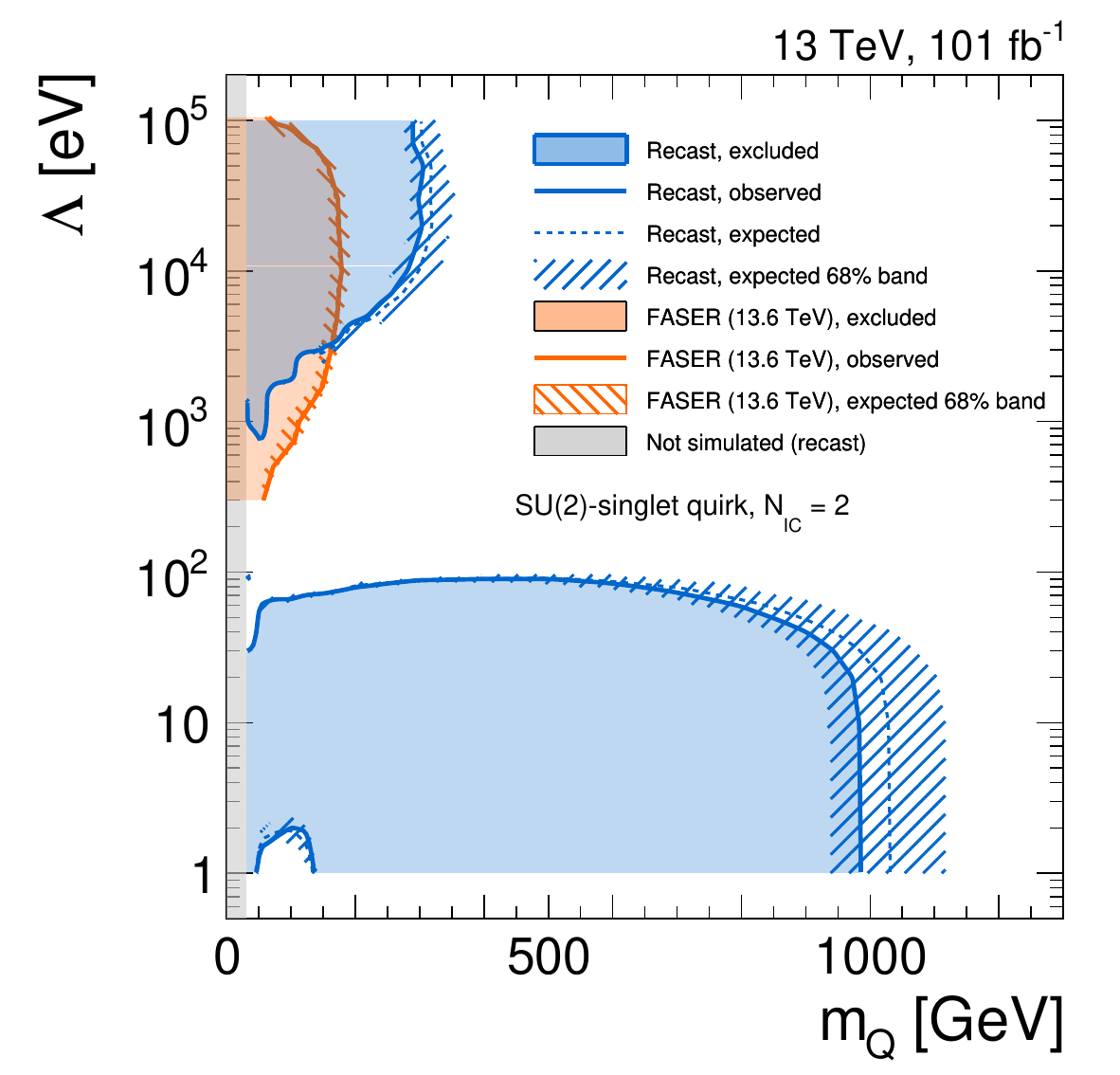}
\caption{Excluded region of an $SU(2)$-singlet quirk with $N_{\mathrm{IC}}=2$ in the $(m_Q,\Lam)$ plane (shaded: excluded), from this reinterpretation of the CMS search at $\sqrt{s}=13$\,TeV (blue) and from the FASER search at $\sqrt{s}=13.6$\,TeV~\cite{FASER:quirksearch} (orange): observed (solid lines) and expected (dashed) with its 68\% band (hatched). The gray region is not simulated in this work.}
\label{fig:faser}
\end{figure}

For a direct comparison with the FASER search~\cite{FASER:quirksearch}, we repeat the analysis for an $SU(2)$-singlet quirk with hypercharge $-1$ and $N_{\mathrm{IC}}=2$, the benchmark of that search. Its couplings to the photon and the $Z$ boson are purely vector, the same as those of the $\taup$ model of the CMS search, so it is generated with the generator settings of the $\taup$ and simulated and analyzed in the same way as the doublet. FASER generated the same Drell-Yan process at LO with MadGraph5\_aMC@NLO~\cite{Alwall:2014hca} interfaced to \textsc{Pythia}~8 (with an LHCb tune and a forward-$Z$ normalization correction) at $\sqrt{s}=13.6$\,TeV, while we use the LO \textsc{Pythia}~8 matrix element at 13\,TeV; both are LO with $N_{\mathrm{IC}}=2$. Figure~\ref{fig:faser} compares the excluded masses. From 4\,keV on, where the pair is reconstructed as one merged track, the CMS search excludes the singlet quirk up to \mSingletTenK\,GeV at $\Lam=10$\,keV and \mSingletThirtyK\,GeV at 30\,keV, beyond the FASER exclusion of 100 to 175\,GeV in this range. From 1 to 3\,keV the two searches overlap: FASER excludes masses up to 110 to 160\,GeV, this recast up to 50\,GeV at 1 to 1.5\,keV and 145 to 155\,GeV at 2 to 3\,keV. From 300 to 600\,eV, where the quirk pair is too wide to be reconstructed as a track, FASER provides the only constraint. Quirk masses below $m_Z/2$ are in any case excluded by the $Z$ width measured at LEP~\cite{ALEPH:2005ab}; we simulate masses down to 30\,GeV only to cover the mass range of FASER. Below 100\,eV, outside the range covered by FASER, the CMS search excludes the singlet up to \mSingletOne\,GeV at $\Lam=1$\,eV and \mSingletSeventy\,GeV at 70\,eV.

\section{Discussion}
\label{sec:discussion}

For the $SU(2)$-singlet quirk with two infracolors and the same definition of $\Lam$, Ref.~\cite{Farina:2017ywr} excluded masses up to 650\,GeV at 1\,eV and 150\,GeV at 100\,eV with 12.9\,fb$^{-1}$ of 2016 data, a looser selection ($p_{\mathrm{T}}>65$\,GeV, $|\eta|<2.1$), and a track model with 30\,$\mu$m hit resolution and without multiple scattering or energy loss in matter. With 101\,fb$^{-1}$ the exclusion at $\Lam=1$\,eV extends to \mSingletOne\,GeV.

The main limitations of this reinterpretation are the following. The trigger efficiency of a bound pair is taken from the efficiency maps of a single slow charged particle, evaluated at the effective velocity of the pair to the muon stations. The ionization response is a parametrization tuned on the published heavy charged lepton likelihood, not a simulation of the CMS silicon readout; saturation of the strip and pixel readout, which matters for charges above $1e$, is not modeled, and a bound pair deposits the ionization of a charge $2e$ only in the sense of the total energy loss of two unit charges along the same path. The isolation uses the generator-level event without pileup, a small effect for a heavy charged lepton according to the published efficiency table. Photon-initiated production, which adds to the DY cross section at high mass, is not included, so the limits are conservative in this respect, as they are with respect to higher-order corrections to the DY cross section.

The approach generalizes to other signatures in which the propagation through the detector, rather than the production, is non-standard: the same chain of a standalone Geant4 simulation in a parametric detector, an emulation calibrated on a benchmark of the original search, and the published likelihoods can be applied whenever a search publishes its statistical model and the signal templates of at least one benchmark.

\section{Summary}
\label{sec:summary}

We have reinterpreted the CMS search for heavy stable charged particles with large ionization energy loss for Drell-Yan produced quirks, with a standalone Geant4 simulation of the quirk dynamics, an emulation of the selection that reproduces the published heavy charged lepton limits within \closureIonRange, and the public likelihoods of the search. For an $SU(2)$-doublet quirk with one (two) infracolors, masses are excluded up to \mObsOne\ (\mObsNicTwoOne)\,GeV at $\Lam=1$\,eV, \mObsSeventy\ (\mObsNicTwoSeventy)\,GeV at 70\,eV, and \mObsTenK\ to \mObsThirtyK\ (\mObsNicTwoTenK\ to \mObsNicTwoThirtyK)\,GeV for $\Lam$ of 10 to 30\,keV, where the pair moves as one merged, doubly ionizing track, while for $\Lam$ from 100 to 800\,eV the search has no sensitivity. For the $SU(2)$-singlet benchmark of FASER, the reinterpretation alone covers $\Lam$ below 100\,eV, up to \mSingletOne\,GeV; above 4\,keV it excludes up to \mSingletTenK\ to \mSingletThirtyK\,GeV, beyond the 100 to 175\,GeV of FASER, while FASER gives the stronger constraint around 1\,keV and the only one from 300 to 600\,eV.

\appendix
\section{Code and data availability}
\label{sec:code}

The quirk simulation (\texttt{quirkcore} and \texttt{quirksim}), the emulation of the CMS selection, the statistical analysis, and the scripts that produce every figure of this paper are available at \url{https://github.com/tvami/QuirkPheno}. The public likelihoods and efficiency maps of the CMS search are taken from HEPData~\cite{CMS:EXO18002:hepdata} without modification.

\section*{Acknowledgments}

We would like to thank Joe Incandela for useful discussions. We used the UCSB computational facilities administered by the Center for Scientific Computing at the California NanoSystems Institute and Materials Research Laboratory (an NSF MRSEC; DMR-1720256) and purchased through NSF CNS-1725797. UCSB is supported by the US Department of Energy under grant DE-SC0011702. Support for UCSB is also made possible by the Joe and Pat Yzurdiaga endowed chair in experimental science. We acknowledge the use of Anthropic's Claude Opus 5.5 model for code development and manuscript editing. All outputs were checked by the authors who take full responsibility for the content.

\bibliography{refs}

\end{document}